# Blockchain-Based Spectrum Resource Securitization via Semi-Fungible Token-Lock

Zhixian Zhou
*College of Electronics and Information Engineering*
*Shenzhen University*
Shenzhen, China
zzxsees@163.com

Bin Chen*
*College of Electronics and Information Engineering*
*Shenzhen University*
Shenzhen, China
bchen@szu.edu.cn
*Corresponding author

Zhe Peng
*Department of Industrial and Systems Engineering*
*The Hong Kong Polytechnic University*
Hong Kong, China
Jeffrey-zhe.peng@polyu.edu.hk

Zhiming Liang
*College of Electronics and Information Engineering*
*Shenzhen University*
Shenzhen, China
lzming0212@163.com

Ruijun Wu
*College of Electronics and Information Engineering*
*Shenzhen University*
Shenzhen, China
19892315202@163.com

Chen Sun
*Wireless Network Research Department*
*Research and Development Center, Sony (China) Limited*
Beijing, China
chen.sun@sony.com

Shuo Wang
*Wireless Network Research Department*
*Research and Development Center, Sony (China) Limited*
Beijing, China
shuo.wang@sony.com

***Abstract*— As 6G networks evolve, spectrum assets require flexible, dynamic, and efficient utilization, motivating blockchain-based spectrum securitization. Existing approaches based on ERC404-style hybrid token models rely on frequent minting and burning during asset transfers, which disrupt token identity continuity and increase on-chain overhead. This paper proposes the Semi-Fungible Token-Lock (SFT-Lock) method, a lock/unlock–based mechanism that preserves NFT identity and historical traceability while enabling fractional ownership and transferability. By replacing mint/burn operations with deterministic state transitions, SFT-Lock ensures consistent lifecycle representation of spectrum assets and significantly reduces on-chain operations. Based on this mechanism, a modular smart-contract architecture is designed to support spectrum authorization, securitization, and sharing, and a staking mechanism is introduced to enhance asset liquidity. Experimental results on a private Ethereum network demonstrate that, compared with ERC404-style hybrid token models, the proposed method achieves substantial gas savings while maintaining functional correctness and traceability.**

## I. Introduction

Amid surging demand for device access and growing service diversity, efficiently managing and flexibly allocating spectrum resources has become a major challenge in 6G network design [1]. Traditional spectrum resource allocation typically relies on static licensing mechanisms [2]. However, this mechanism faces issues such as spectrum underutilization and difficulties in accessing spectrum [3]. To address these problems, Dynamic Spectrum Access (DSA) has become a key strategy to improve spectrum utilization efficiency. DSA technology offers a more flexible spectrum management approach to spectrum management by allowing secondary users (SU) to utilize idle frequency bands without interfering with primary users (PU), thereby improving spectrum utilization [4]. Many countries and regions are actively exploring spectrum sharing methods based on DSA, such as the CBRS and the LSA [5]. However, DSA faces several challenges in practical applications, including centralized management risks, transaction record tampering, and inefficient authorization processes [6].

To address the aforementioned challenges, an increasing number of studies have begun applying blockchain technology to spectrum management. The core characteristics of blockchain, such as decentralization, immutability, and programmability, offer new solutions for spectrum sharing, trading, and regulation [7]. Blockchain-based smart contracts enable the automation of key processes including spectrum licensing, revenue allocation, thereby enhancing both operational efficiency and system trustworthiness [8]. The feasibility of leveraging blockchain technology for spectrum sharing has been explored in prior studies. For instance, Perera et al. [9] introduced the Spect-NFT framework. Shao et al. [6] proposed an NFT-based dynamic spectrum sharing model for 6G networks. Ye et al. [10] and Liang et al. [11] utilized the ERC4907 standard [12] to mint rentable non-fungible spectrum tokens. However, the solutions mentioned above still lack a mechanism to incentivize users to share spectrum resources.

Zhou et al. [13] proposed an ERC404-style hybrid token models based securitization approach for spectrum resources. By tokenizing spectrum holdings, primary users PUs can reduce asset holding risk, improve liquidity, and strengthen incentives to participate in spectrum sharing. ERC404-style hybrid token models [14] is a community proposed and experimental hybrid token design that combines non-fungible token (NFT) [15] style identity with fungible token (FT) [16] accounting to support fractionalization and tradability. However typical ERC 404 style implementations couple FT transfers with NFT mint and burn operations. When balances cross unit thresholds NFT instances are destroyed and recreated. This prevents any single NFT from persistently representing the same underlying spectrum asset and weakens continuity traceability and state stability. As a result in [13] NFTs effectively encode only the quantity of spectrum held by a PU but they cannot faithfully track the spectrum resource dynamic operational status.

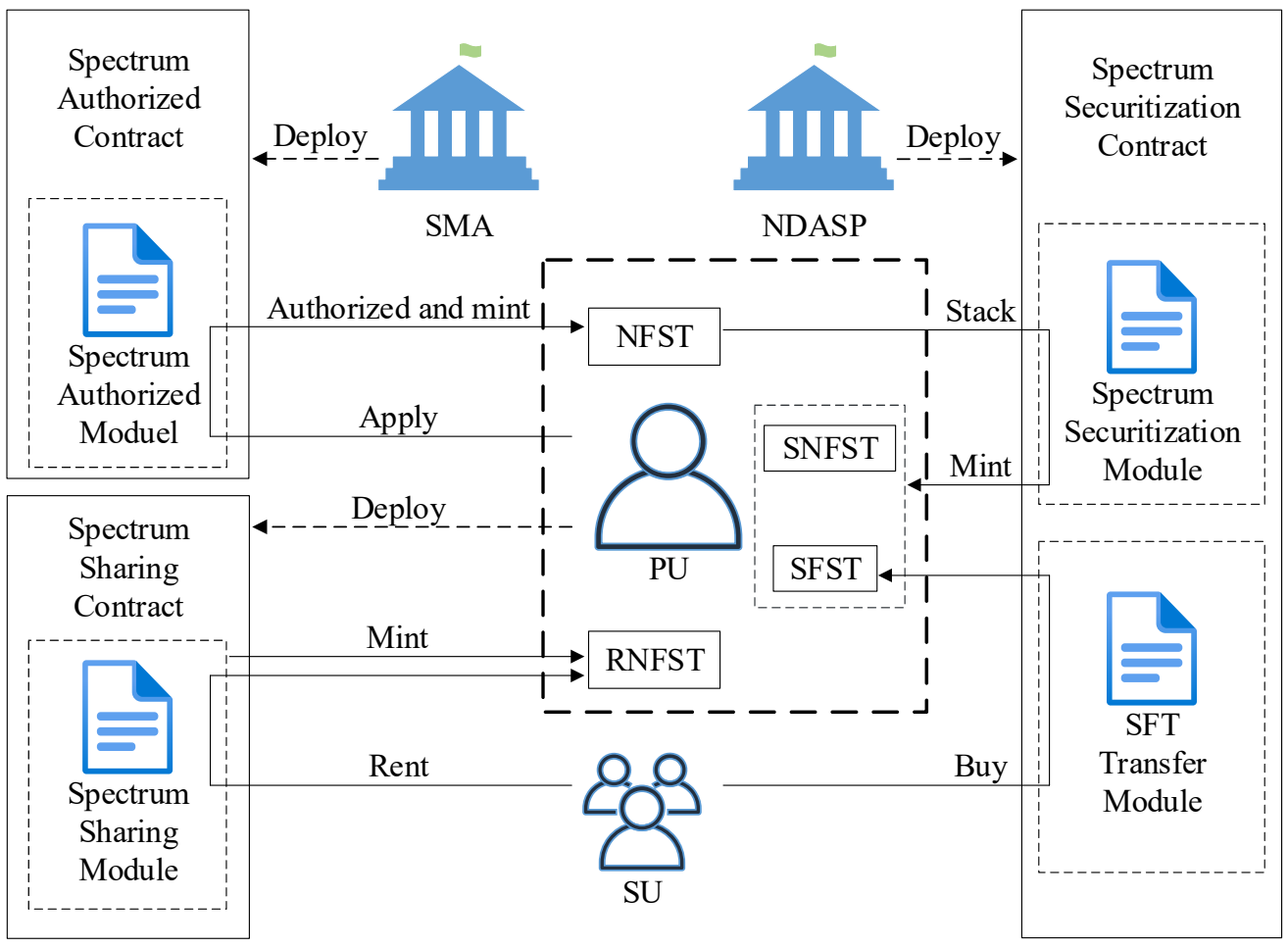

Fig. 2.1. System model for spectrum securitization.

To overcome the structural limitations identified in [13], this paper designs a spectrum securitization system and proposes the SFT-Lock method, which replaces the traditional mint/burn mechanism with a lock/unlock design. Unlike ERC404-style hybrid token models, SFT-Lock does not redefine token standards but introduces a state-transition mechanism that can be layered on hybrid asset models. SFT-Lock eliminates the need for token destruction, thereby preserving the identity and historical continuity of NFTs throughout their lifecycle. It further allows users to customize the locking and unlocking sequence, offering flexible control over asset states. In addition, a staking mechanism is introduced to transform static spectrum resources into tradable securities, effectively unlocking their liquidity and enhancing asset utility.

## II. System Model

This section introduces the system model of the proposed spectrum securitization system. This system consists of three contracts. The Spectrum Authorized Contract is responsible for the initial tokenization of spectrum resources into Non-Fungible Spectrum Tokens (NFSTs). The Spectrum Securitization Contract allows PUs to stake their NFSTs in exchange for Securitized NFSTs (SNFSTs) and Securitized Fungible Spectrum Tokens (SFSTs). The Spectrum Sharing Contract enables the sharing of unused spectrum resources with SUs.

### A. Roles and Contract Interactions

Figure 2.1 illustrates the interaction between smart contracts and the primary roles within the system. This system involves four roles: SMA (Spectrum Management Authority), NDASP (National Digital Asset Securitization Platform), PU and SU, they participate in the system through smart contracts:

*1) SMA:* Responsible for the initial authorization and registration of spectrum resources. It deploys the Spectrum Authorized Contract, which includes:

*a) Spectrum Resource Authorization Module:* Enables SMA to mint spectrum resources into NFSTs for PUs.

*b) Spectrum Resource Reclaim Module:* Enables SMA to reclaim NFST held by PUs.

*2) NDASP:* Manages the securitization of spectrum assets. It deploys the Spectrum Securitization Contract, which includes:

*a) NFST Staking Module:* Enables PUs to stake NFSTs to initiate securitization and mint SNFSTs and SFSTs to PUs.

*b) SFST Transfer Module:* Enables users to transfer SFSTs between each other, and leverages the lock/unlock mechanism of the SFT-Lock method to manage the state transitions of SNFSTs.

*3) PU:* Deploys the Spectrum Sharing Contract to facilitate efficient spectrum utilization. This contract incorporates a Spectrum Leasing Module, which is built upon the ERC4907 standard and enables PUs to mint Rental NFSTs (RNFSTs). SUs can acquire RNFSTs by exchanging SFSTs, thereby gaining temporary access to spectrum resources.

### B. SFT-Lock method

To address the defects of the ERC404 style hybrid token model in spectrum resource securitization, such as the reliance on frequent minting and burning operations that compromise NFT traceability and state consistency, this paper proposes SFT-Lock. SFT-Lock decouples asset divisibility from token identity by introducing an explicit state transition layer. It maintains a set of spectrum NFTs and maps fungible share transfers to deterministic lock or unlock operations. Each spectrum NFT persists with an explicit locked or unlocked state. The specific logic is as follows:

*1) NFT lock:* When a FT is split, the master contract updates a NFT status to "locked".

*2) NFT unlock:* When the PU aggregates a FT, the contract resets a NFT status to "unlocked".

The detailed diagram of the SFT-Lock method is shown in Figure 2.2.

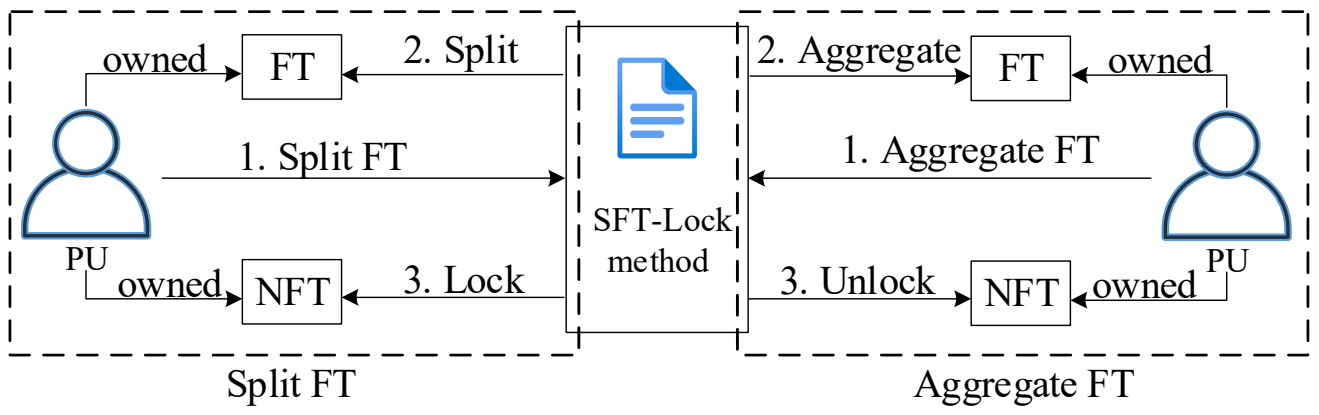

Fig. 2.2. SFT-Lock method.

## III. Smart Contract Design

### *A. Contract Architecture and State Transition*

This system implements blockchain based spectrum securitization using three on chain contracts that separate authorization, securitization, and sharing [1][4][8]. The Spectrum Authorization contract maintains the NFST registry, including spectrum metadata such as channel and location and basic status flags, and exposes mintNFST and reclaimNFST. The Spectrum Securitization contract escrows NFSTs and mints securitized representations. It records an NFST ID to original owner mapping and maintains each PU inventory of SNFST IDs and SFST balances. The stakeNFST function transfers custody of the NFST to escrow and mints SNFST and SFST to the PU. The Spectrum Sharing contract provides time bounded usage rights using the rentable NFT workflow [10]-[12], minting rental NFTs that can be rented by SUs while preserving the underlying securitization state.

After securitization, each spectrum unit is represented by an SNFST with persistent NFT identity [15], while fractional ownership is represented by SFST balances under ERC20 accounting [16]. Each SNFST corresponds to one spectrum unit. Let share be the integer number of whole units implied by an address SFST balance, where integer division rounds down. After any SFST transfer, SNFST state changes are triggered only when the integer share of the PU address changes. If the PU share increases, the system unlocks the same number of SNFST units. If the PU share decreases, the system locks the same number of SNFST units. Transfers between SUs do not change SNFST states unless they change the PU integer share. Optional lockOrder and unlockOrder lists specify which SNFST IDs transition first. For sharing, a PU mints a rental NFT and an SU rents it under the ERC4907 workflow, while the underlying securitization ownership state remains unchanged [10]-[12].

### *B. Spectrum Authorization Module*

The spectrum authorization module initializes on chain spectrum assets through two procedures, mintNFST() and reclaimNFST(). When a PU is approved for spectrum usage, SMA calls mintNFST() to create an NFST that records the allocated spectrum metadata and serves as the digital representation of the spectrum asset. The procedure is specified in Algorithm 1. When the allocation is revoked or expires, SMA calls reclaimNFST() to reclaim the corresponding NFST. The procedure is specified in Algorithm 2.

**Algorithm 1:** mintNFST

```
Input: toAddress, channel, location
Increment global tokenIdCounter
Set newTokenId = tokenIdCounter
Create a new NFST entry:
    tokenId = newTokenId
    owner = toAddress
    channel = channel
    location = location
Mark (channel, location) as uploaded
Append newTokenId to:
    Global minted NFST list
    List of NFSTs owned by toAddress
If toAddress is not already marked as a PU then
    Mark toAddress as a PU
```

**Algorithm 2:** reclaimNFST

```
Input: tokenId
Retrieve NFST info associated with tokenId
Mark (channel, location) as not uploaded
Remove tokenId from:
    Global minted NFST list
    List of NFSTs owned by contract owner
Mark the NFST as reclaimed
If contract owner no longer holds any NFSTs then
    Mark contract owner as not a PU
```

### *C. Spectrum Securitization Module*

**Algorithm 3:** stakeNFST

```
Input: PUAddress, tokenIdOfNFST
Retrieve NFST metadata from SMA:
    owner, channel, location ← getNFSTInfo(tokenIdOfNFST)
Require: PU is the current owner of the NFST
Record original owner:
    originOwnerOfNFST[tokenIdOfNFST] ← PU
Call the function of SMA's contract to update stake info:
    Set NFST's staking status to true
    Assign current contract as the holder of the NFST
Mint a SNFST for PU, using tokenIdOfNFST, channel, location
Mint SFST for PU
```

PUs seeking to securitize their spectrum assets may invoke the stakeNFST() function to stake an NFST into the Spectrum Securitization Contract. Upon successful staking, the contract mints a corresponding SNFST and SFST for the PU, which together represent the tokenized form of the underlying spectrum resource. Ownership of the original NFST is escrowed in the securitization contract, while usage and economic rights are represented by the SNFST/SFST pair. The logic of the stakeNFST() function is detailed in Algorithm 3.

### *D. SFST Transfer Module*

This module supports the core operations needed to lock and unlock spectrum unit tokens, allow a PU to specify a preferred transition order, and carry out SFST transfers that trigger SNFST state updates when the PU integer share changes. The lock and unlock procedure is specified in Algorithm 4 and 5.

**Algorithm 4:** lockSNFST

```
Input: PUAddress, tokenId, isOrder
Retrieve SNFST info associated with tokenId
Mark SNFST as locked
Remove tokenId from:
    unlockedSNFST[PU]
Add tokenId to:
    lockedSNFST[PU]
If isOrder is true then
    Remove tokenId from lockOrder[PU]
```

**Algorithm 5:** unlockSNFST

**Input:** PUAddress, tokenId, isOrder

```
Retrieve SNFST info associated with tokenId
Mark SNFST as unlocked
Remove tokenId from:
    lockedSNFST[PU]
Add tokenId to:
    unlockedSNFST[PU]
If isOrder is true then
    Remove tokenId from unlockOrder[PU]
```

The system also supports customizing the locking and unlocking order of SNFSTs. This design enables policy-aware state transitions when securitized spectrum assets exhibit heterogeneous attributes such as expiration time, geographic scope, or regulatory priority. The logic of setLockOrder() function and setUnlockOrder() function is shown in Algorithm 6 and Algorithm 7.

**Algorithm 6:** setLockOrder

**Input:** tokenIdList

```
For each tokenId in tokenIdList do
    Retrieve SNFST info
    Require: Caller is the owner of the SNFST
    Require: SNFST is not currently locked
Set lockOrder[caller] to tokenIdList
```

**Algorithm 7:** setUnlockOrder

**Input:** tokenIdList

```
For each tokenId in tokenIdList do
    Retrieve SNFST info
    Require: Caller is the owner of the SNFST
    Require: SNFST is currently locked
Set unlockOrder[caller] to tokenIdList
```

The transfer() function implements the lock/unlock mechanism defined in the SFT-Lock method and facilitates the transfer of SFSTs between users. Because each SNFST corresponds to a fixed unit of spectrum capacity, changes in SFST balances deterministically imply the number of SNFSTs that must transition between locked and unlocked states. The transfer logic deterministically maps SFST balance changes to SNFST state transitions, ensuring a one-to-one correspondence between fungible shares and locked spectrum units. The logic of the transfer() method is illustrated in Algorithm 8.

**Algorithm 8:** transfer

**Input:** from, to, PU, amount

```
unit ← 1 × 10^18
from_OldShare ← floor(userShares[PU][from] / unit)
to_OldShare ← floor(userShares[PU][to] / unit)
Call transferSFST(from, to, PU, amount)
from_CurrentShare ← floor(userShares[PU][from] / unit)
to_CurrentShare ← floor(userShares[PU][to] / unit)
If from == PU then
    lockNum ← max(0, from_OldShare - from_CurrentShare)
    orderLockNum ← min(lockNum, lockOrder[PU].length)
    For i from 1 to orderLockNum do
        lockSNFST(PU lockOrder[PU][0] true)
    For i from 1 to lockNum - orderLockNum do
        lockSNFST(PU unlockedSNFST[PU][0] false)
If to == PU then
    unlockNum ← max(0, to_CurrentShare - to_OldShare)
    orderUnlockNum←min(unlockNum, unlockOrder[PU].length)
    For i from 1 to orderUnlockNum do
        unlockSNFST(PU unlockOrder[PU][0] true)
    For i from 1 to unlockNum - orderUnlockNum do
        unlockSNFST(PU lockedSNFST[PU][0] false)
```

## *E. Spectrum Sharing Module*

Spectrum Sharing Module implements spectrum resource leasing function based on ERC4907 standard, enabling short-term renting of spectrum resources. The minting and renting methods for RNFST can refer to reference [10]-[12].

# IV. Experimental Results and Analysis

To simulate real on-chain interactions, Ganache private blockchain was chosen as the underlying operating network. Contract development and deployment were conducted using Solidity 0.8.x. Table I lists the basic information of each user and the contract addresses of Spectrum Authorized Contract and Spectrum Securitization Contract.

TABLE I. Basic Information and Description of Each User

| Role | Address | Description |
|---|---|---|
| SMA | 0x50C…BCB | Allocate and reclaim spectrum resources and deploy Spectrum Authorized Contract |
| NDASP | 0x302…03a | Responsible for spectrum securitization and deployment of Spectrum Securitization Contract |
| PU | 0x0aa…f60 | Receive spectrum SNFSTs and SFSTs, which can be securitized and shared with SUs |
| SU | 0x408…F9F | Buy SFSTs and rent RNFST using ETH |

## *A. Functional Verification*

In order to verify the effectiveness of the securitization system based on the SFT-Lock method proposed in this paper, a series of experiments were designed and carried out.

```
{
        "from": "0x748986d49e1504df069b7075edfe0b1fe8a7b646",
        "topic": "0x1279982e0ba8e0cbe86f8ba2c77157e9ae8e0e74ed8a36344e88c06eba3ed8cd",
        "event": "MINT_NFST",
        "args": {
                "0": "0x0000000000000000000000000000000000000000",
                "1": "0x0aa7652B45d957B9d2dE60AFbbD90b2DaD3d1f60",
                "2": "1",
                "_from": "0x0000000000000000000000000000000000000000",
                "_to": "0x0aa7652B45d957B9d2dE60AFbbD90b2DaD3d1f60",
                "_tokenId": "1"
        }
}
```

Fig. 4.1. Results of minting NFST.

SMA mints NFST by inputting the metadata of spectrum resources. Figure 4.1 shows the smart contract execution result of SMA minting NFST for the PU. After NFST is successfully minted, the MINT_NFST event is triggered, recording that the contract successfully minted spectrum NFST numbered 1 for the PU. In the event parameters, _from is an all-zero address, indicating that the NFT is newly minted rather than transferred. _to indicates the recipient address, and _tokenId identifies the unique number of the NFST.

When the right to use spectrum resources needs to be reclaimed due to expiration, SMA can initiate a mandatory recovery operation on the NFST held by the PU to ensure the compliance of resource use. As shown in Figure 4.2, when SMA reclaims NFST, the RECLAIM_NFST event is triggered, indicating that the NFST numbered 2 has been reclaimed. In the parameter, the _from field indicates that the owner of the NFST is changed from the PU to SMA, and the reclaim operation is performed by SMA.

```
{
        "from": "0x748986d49e1504df069b7075edfe0b1fe8a7b646",
        "topic": "0x016c0a85e2495dd981afb4ce5a73059e035b4d946453365a08bfccf0c7b102cd",
        "event": "RECLAIM_NFST",
        "args": {
                "0": "0x50C0720D772D21017dD0bD4D1Cb1357B3dc59BCB",
                "1": "3",
                "_from": "0x50C0720D772D21017dD0bD4D1Cb1357B3dc59BCB",
                "_tokenId": "3"
        }
}
```

Fig. 4.2. Results of SMA reclaim NFST.

```
{
        "from": "0xd30c2759bae44819d4817568649664750a78b10f",
        "topic": "0x2512fffbfab4a0e58450e89290d5eca31e2ad1798f0635c73777321f3d77dc4c",
        "event": "TRANSFER_SNFST",
        "args": {
                "0": "0x0000000000000000000000000000000000000000",
                "1": "0x0aa7652B45d957B9d2dE60AFbbD90b2DaD3d1f60",
                "2": "1",
                "_from": "0x0000000000000000000000000000000000000000",
                "_to": "0x0aa7652B45d957B9d2dE60AFbbD90b2DaD3d1f60",
                "_tokenId": "1"
        }
},
{
        "from": "0xd30c2759bae44819d4817568649664750a78b10f",
        "topic": "0x35f5fbba9b3a7f097facb1f22fba98c1b1fcbdeeefed8d7e1f27dc2f5cdbfc55",
        "event": "TRANDFER_SFST",
        "args": {
                "0": "0x0000000000000000000000000000000000000000",
                "1": "0x0aa7652B45d957B9d2dE60AFbbD90b2DaD3d1f60",
                "2": "0x0000000000000000000000000000000000000000",
                "3": "1000000000000000000",
                "_from": "0x0000000000000000000000000000000000000000",
                "_to": "0x0aa7652B45d957B9d2dE60AFbbD90b2DaD3d1f60",
                "_primaryUser": "0x0000000000000000000000000000000000000000",
                "_amount": "1000000000000000000"
        }
},
```

Fig. 4.3. Spectrum resource securitization.

When the PU needs to securitize the spectrum resources it holds, it can stake NFST to the Spectrum Securitization Contract. As shown in Figure 4.3, the following events will be triggered during the securitization process:

*1) TRANSFER_SNFST:* Records that the contract successfully minted SNFST numbered 1 for the PU. In the event parameters, _from is an all-zero address, indicating that the NFT is newly minted. _to indicates the recipient's address, and _tokenId identifies the unique number of the SNFST.

*2) TRANSFER_SFST:* Records that the contract mints the SFST corresponding to SNFST numbered 1 to the PU. The parameters include _from, _to, _primaryUser, and _amount (in wei, 1 ETH = $1 \times 10^{18}$ wei).

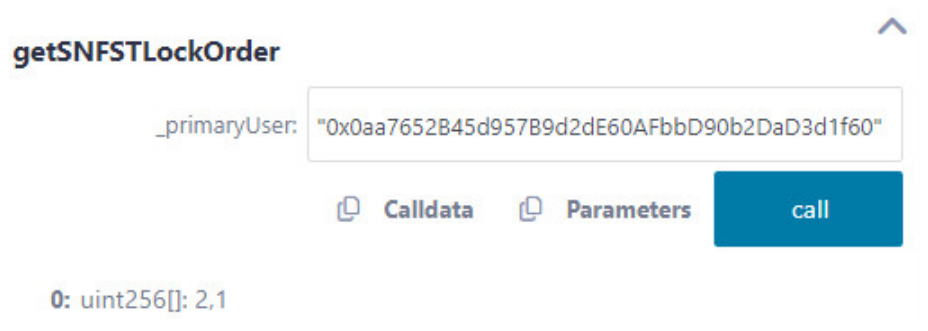

Fig. 4.4. SNFST lock order.

The PU securitizes $NFST_1$ and $NFST_2$ sequentially, obtaining $SNFST_1$ and $SNFST_2$. Suppose that, due to business requirements, the PU must lock $SNFST_2$ before $SNFST_1$. By default, the locking order is $SNFST_1 \rightarrow SNFST_2$. Therefore, the PU modifies the priority locking order. As illustrated in Figure 4.4, the PU sets the locking order to $SNFST_2 \rightarrow SNFST_1$, ensuring that $SNFST_2$ is locked first.

```
{
        "from": "0xd30c2759bae44819d4817568649664750a78b10f",
        "topic": "0x35f5fbba9b3a7f097facb1f22fba98c1b1fcbdeeefed8d7e1f27dc2f5cdbfc55",
        "event": "TRANDFER_SFST",
        "args": {
                "0": "0x0aa7652B45d957B9d2dE60AFbbD90b2DaD3d1f60",
                "1": "0x408DD44B2c2Ebfdd0f9B66A448eEa7293B3c1F9F",
                "2": "0x0aa7652B45d957B9d2dE60AFbbD90b2DaD3d1f60",
                "3": "300000000000000000",
                "_from": "0x0aa7652B45d957B9d2dE60AFbbD90b2DaD3d1f60",
                "_to": "0x408DD44B2c2Ebfdd0f9B66A448eEa7293B3c1F9F",
                "_primaryUser": "0x0aa7652B45d957B9d2dE60AFbbD90b2DaD3d1f60",
                "_amount": "300000000000000000"
        }
},
{
        "from": "0xd30c2759bae44819d4817568649664750a78b10f",
        "topic": "0xc84ade9ebbaa67935f2681d0fba98887fd0f62279b647a933fb58cd8cacbaad0",
        "event": "LOCK_SNFST",
        "args": {
                "0": "0x0aa7652B45d957B9d2dE60AFbbD90b2DaD3d1f60",
                "1": "2",
                "_primaryUser": "0x0aa7652B45d957B9d2dE60AFbbD90b2DaD3d1f60",
                "_tokenId": "2"
        }
},
```

Fig. 4.5. PU transfer SFSTs to SU

```
{
        "from": "0xd30c2759bae44819d4817568649664750a78b10f",
        "topic": "0x35f5fbba9b3a7f097facb1f22fba98c1b1fcbdeeefed8d7e1f27dc2f5cdbfc55",
        "event": "TRANDFER_SFST",
        "args": {
                "0": "0x408DD44B2c2Ebfdd0f9B66A448eEa7293B3c1F9F",
                "1": "0x0aa7652B45d957B9d2dE60AFbbD90b2DaD3d1f60",
                "2": "0x0aa7652B45d957B9d2dE60AFbbD90b2DaD3d1f60",
                "3": "300000000000000000",
                "_from": "0x408DD44B2c2Ebfdd0f9B66A448eEa7293B3c1F9F",
                "_to": "0x0aa7652B45d957B9d2dE60AFbbD90b2DaD3d1f60",
                "_primaryUser": "0x0aa7652B45d957B9d2dE60AFbbD90b2DaD3d1f60",
                "_amount": "300000000000000000"
        }
},
{
        "from": "0xd30c2759bae44819d4817568649664750a78b10f",
        "topic": "0xe6ba66b0e6eac7a2a7ebf7c89e8ba1d431b60fad6faf265be810086eeb8bf1a4",
        "event": "UNLOCK_SNFST",
        "args": {
                "0": "0x0aa7652B45d957B9d2dE60AFbbD90b2DaD3d1f60",
                "1": "2",
                "_primaryUser": "0x0aa7652B45d957B9d2dE60AFbbD90b2DaD3d1f60",
                "_tokenId": "2"
        }
},
```

Fig. 4.6. SU transfer SFSTs to PU.

Subsequently, the PU calls the transfer() method to transfer 0.3 SFSTs to the SU. The execution result of the smart contract is shown in Figure 4.5. The LOCK_SNFST event indicates that $SNFST_2$ is locked. In the parameters, _primaryUser is still the PU, and _tokenId is the number of the locked SNFST. These results confirm that spectrum securitization and subsequent transfers do not create or destroy NFTs, but only induce state transitions, thereby preserving NFT identity continuity.

TABLE II. SFST TRANSFER INFORMATION AND $SNFST_2$'S STATUS

| User | Number of SFST before transfer | Number of SFST after transfer | $SNFST_2$ status |
|---|---|---|---|
| PU | 2 | 1.7 | lock |
| SU | 0 | 0.3 | N/A |
| PU | 1.7 | 2 | unlock |
| SU | 0.3 | 0 | N/A |

In order to verify the unlocking function of SNFST, the SU transfers 0.3 SFSTs to the PU. The execution result of the smart contract is shown in Figure 4.6. The UNLOCK_SNFST event represents the triggering of the SNFST unlocking event. In the parameters, "_primaryUser" is the original holder address, and "_tokenId" is the ID of the unlocked SNFST. Table II shows the changes in SFSTs balances of the PU and the SU and the changes in $SNFST_2$ status during the two transfers.

### *B. Performance comparison of SFT-Lock method and ERC404-style hybrid token model*

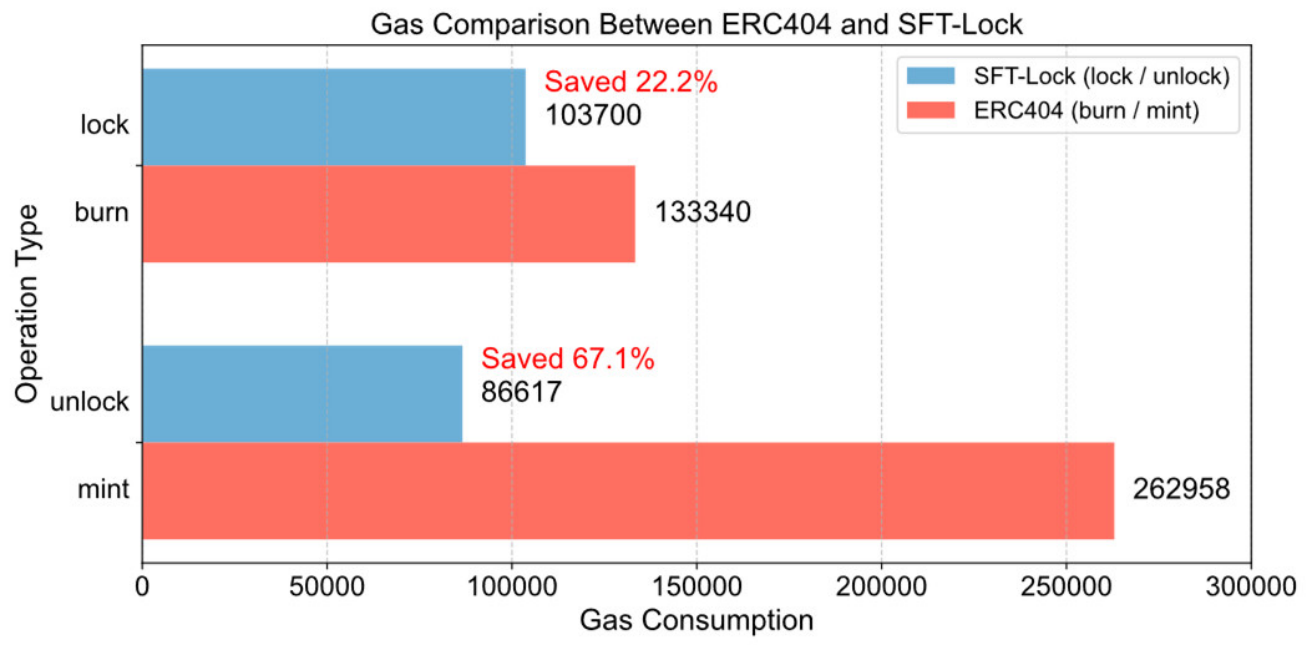

Fig. 4.8. Gas consumption comparison

To compare the gas consumption of an ERC-404-style mint/burn hybrid model and the proposed SFT-Lock method, we implemented equivalent spectrum securitization modules for both approaches. Both implementations were evaluated under identical contract logic and compiler settings, differing only in the use of mint/burn operations versus lock/unlock state transitions. In Fig. 4.8, under the ERC404-style hybrid token model, minting and burning operations consume 262,958 and 133,340 gas, respectively. In contrast, the SFT-Lock method replaces these operations with lightweight unlock and lock state transitions, which consume 86,617 and 103,700 gas, respectively. This corresponds to a 67.1% reduction in gas consumption for unlock compared to minting, and a 22.2% reduction for lock compared to burning. Overall, the results demonstrate that SFT-Lock significantly reduces on-chain gas consumption and operational complexity relative to the ERC404-style hybrid token model. This improvement arises from the fact that SFT-Lock avoids token creation and destruction, relying instead on constant-time state updates that eliminate the storage allocation and deallocation overhead inherent to minting and burning operations.

## V. Conclusion

This paper presents a blockchain based spectrum resource securitization system and introduces the SFT-Lock method. SFT-Lock replaces the mint and burn operations used in ERC 404 style designs with a lock and unlock mechanism, which avoids the continuity and traceability issues caused by repeated NFT recreation. This design better fits spectrum securitization and reduces gas consumption relative to ERC 404 style mint and burn approaches. The system also provides a user-defined lock/unlock priority feature to enable fine-grained asset management. In addition, a staking mechanism is introduced to convert static spectrum assets into tradable securities, thereby enhancing their liquidity.

## VI. Acknowledgment

This work is supported in part by the Foundation of Guangdong Province Graduate Education Innovation Program Project[2024JGXM_163], Shenzhen University High-Level University Construction Phase III – Human and Social Sciences Team Project for Enhancing Youth Innovation[24QNCG06].